# Circumstantial-Evidence-Based Judgment for Software Effort Estimation

Zheng Li[1,2], Liam O'Brien[3,2] and He Zhang[1,4]

[1] NICTA, Sydney, Australia
{Zheng.Li, He.Zhang}@nicta.com.au
[2] School of CS, ANU, Canberra, Australia
[3] CSIRO, Canberra, Australia
Liam.OBrien@csiro.au
[4] School of CSE, UNSW, Sydney, Australia

**Abstract.** Expert judgment for software effort estimation is oriented toward direct evidences that refer to actual effort of similar projects or activities through experts' experiences. However, the availability of direct evidences implies the requirement of suitable experts together with past data. The circumstantial-evidence-based judgment proposed in this paper focuses on the development experiences deposited in human knowledge, and can then be used to qualitatively estimate implementation effort of different proposals of a new project by rational inference. To demonstrate the process of circumstantial-evidence-based judgment, this paper adopts propositional learning theory based diagnostic reasoning to infer and compare different effort estimates when implementing a Web service composition project with some different techniques and contexts. The exemplar shows our proposed work can help determine effort tradeoff before project implementation. Overall, circumstantial-evidence-based judgment is not an alternative but complementary to expert judgment so as to facilitate and improve software effort estimation.

## 1 Introduction

Mathematical effort estimation models have been well documented in academia for many years, while the pervasive estimation method in industry is still based on expert judgment [1]. One possible reason is that the mental processes software professionals use to unfold estimation are more closely related to a case-based reasoning (CBR) approach than a regression-based model [2]. However, expert judgment considerably depends on experts' availability and experience, and experts' knowledge is hardly accessed by others [3]. Therefore, expert opinion may be not reliable if it is not supported by objective or scientific evidences. To reduce the possible bias and uncertainty that happens in expert judgment, practical guidelines claim that estimation experts should be selected based on their experience from similar projects [4]. Following the practical guidelines, unfortunately, the expert judgment approach could still be infeasible if the experts or the past data are not available.

Basically, expert judgment-based software effort estimation must comply with a golden rule: the expert judgment should always require justification rather than gut feelings [4]. Inspired by Evidence-Based Software Engineering (EBSE) that is "to provide the means by which current best evidence from research can be integrated with practical experience and human values in the decision making process regarding the development and maintenance of software" [5], we can re-consider the justification of expert judgment from an evidence-based perspective. According to the classification of evidence [6], the results of the CBR-based mental processes in traditional expert judgment can be regarded as direct evidence: experts act as witnesses and adduce previous cases for the current one. Considering the aforementioned limitation of direct evidence collection – the requirement of availability of experts with experience from similar projects, this paper proposes circumstantial-evidence-based judgment for software effort estimation. Benefiting from existing software development experiences as circumstantial evidence, we can use diagnostic reasoning to qualitatively infer different implementation effort of different proposals of a new project. As a result, circumstantial-evidence-based judgment can be used to facilitate and improve the final quantitative effort estimation for new software projects.

This paper is organized as follows. Section 2 makes a comparison between direct and circumstantial evidences for effort judgment. Section 3 introduces the method and inference procedure that can support circumstantial-evidence-based effort judgment. Section 4 takes Web service composition project as sample to demonstrate the process and result of effort judgment with circumstantial evidences. The conclusion is drawn, and some future research work is considered in Section 5.

## 2 Direct vs. Circumstantial Evidence for Effort Judgment

As an analogue of similar forensic scenarios, the traditional CBR-based expert judgment can be viewed as using direct evidence to estimate implementation effort of a software project. In forensic science, as the name suggests, direct evidence is evidence that proves a fact without requiring inference or presumption [6]. In other words, direct evidence immediately and precisely establishes a bridge between judge and fact. An example of direct evidence could be a witness's observation or personal knowledge of a certain fact. The defendant involved in the past fact is the exact one involved in the judgment. In the context of effort estimation, different from law, the new project to be judged is obviously none of the past ones. Nevertheless, to some extent, experts inevitably view the similar projects as the same one when doing effort estimation based on their experiences. For example, as suggested by Jørgensen [4], the actual effort of similar projects or similar activities in other projects will be referred to as justifications for expert judgment for new project. Therefore, it is reasonable to consider that in traditional expert judgment experts use their observation on past projects as direct evidences to estimate effort of new project.

Contrasted with direct evidence, circumstantial evidence does not prove fact in a straightforward sense, while it requires the intervening or additional evidence inference to confirm the fact. In forensic science, the most obvious difference between direct and circumstantial evidence is that "direct evidence is a verbal representation of a crime itself, whereas circumstantial evidence is an abstract statement about the connection between the defendant and an incriminating physical trace of the crime" [7]. Usually, circumstantial evidence is not sufficient, but increases the probability of the defendant's guilt, for example blood or fingerprints. Similarly, unlike the direct evidence in expert judgment that directly gives the estimated effort, circumstantial evidence for effort judgment must be effort-related abstract statements. Suppose each finished software project deposits some development experience in the human knowledge, similar projects or similar activities should have similar development experiences. Different experiences can then be abstracted into different assertions as scattered fingerprints of existing software projects. As such, different from human beings, similar software projects or development activities may share the same fingerprints.

To sum up, when estimating effort for a new project, similar projects' or activities' actual effort can be viewed as direct evidence, while existing development experiences can be considered as circumstantial evidence. In forensic science, both direct and circumstantial evidences are used to draw categorical, yes-no type, conclusions. In the context of software effort estimation, direct evidence brings quantitative effort estimate for a particular project proposal, while circumstantial evidence can give qualitative comparison between effort estimates of different development proposals.

## 3  Effort Judgment with Circumstantial Evidence

### 3.1  Collecting Circumstantial Evidences

When it comes to collecting evidences for effort estimation, direct evidence collection is to gather detailed software project data, while circumstantial evidence collection is to gather generic software development experiences. Compared with the straightforward process of direct evidence collection, searching and identifying circumstantial evidences could be more complicated. We propose to use a systematical method that is to apply systematic literature review (SLR) in the evidence space formed by all the effort factors. SLR is the main methodology applied for EBSE, which can be naturally used to collect and justify different effort-related hypothesis aiming at different effort factors. The justified effort-related hypothesis can then be used as circumstantial evidences for effort judgment. As for the effort factors, we can directly borrow ideas from existing effort estimation work, such as parametric estimation models. For example, 15 effort multipliers like Product Complexity and Programmers' Capability are employed in COCOMO [20], and each of them can be viewed as an effort factor towards which we may identify a corresponding circumstantial evidence through SLR.

### 3.2  Utilizing Circumstantial Evidences

As specified previously, circumstantial evidences cannot be used without rational inference to proven facts. The rational inference can be realized as a cascaded process of diagnostic reasoning. A possible guideline for using circumstantial evidences to do diagnostic reasoning is the theory of propositional learning (TPL) [15]. TPL is originally used for belief revision, which comprises three elements: (1) the association between a possible clue and a possible cause; (2) the forward implication from the actual cause to the possible clue; (3) the backward implication from the clue to the possible cause. The clues are circumstantial evidences like fingerprints, while the causes are suspects' actions by which the fingerprints are left. When implementing diagnostic reasoning with TPL, the inference process of diagnostic reasoning can then be established through the linkage of aforementioned elements, as illustrated in Fig. 1.

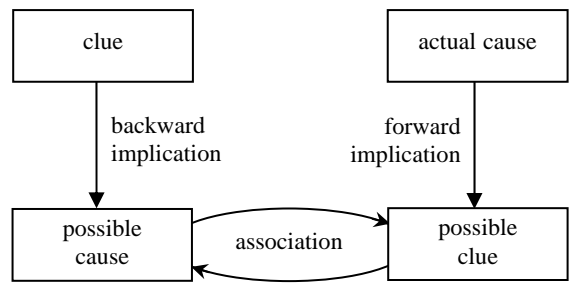

**Fig. 1.** The inference process of diagnostic reasoning.

For software effort judgment, the first-hand clues are existing software projects, while the actual cause is the requirement of a new project. When doing backward implication, benefiting from the techniques of EBSE [5], the original clues can be collected and used to extract effort-related assertions. Note that, different from the work in [2] that collects the actual effort of similar projects, EBSE used for circumstantial-evidence-based judgment focuses on the generic relationships between effort and different development actions, namely development experiences. When doing forward implication, on the other hand, the profile of a concrete project will be explored to identify possible development actions. The identified possible development actions can be used to build an association between previous projects and the current one to further facilitate effort judgment. In general, the association is built by a cascaded inference. In a cascaded inference, the conclusion of one inference acts as a premise for the subsequent inference, while the final conclusion will be the qualitatively estimated effort.

In practice, there is usually a set of circumstantial evidences for one effort judgment task. These circumstantial evidences can be either consistent with or contrary to each other. Here we define different evidences are *consistent* when the same conclusion can be drawn in an effort judgment, or *contrary* when different conclusions are drawn in the judgment. Generally, consistent circumstantial evidences can help confirm and reinforce the same conclusion. For contrary circumstantial evidences, we would have to give further judgment based on the amount and weight of different standpoints, which is similar to the reality in a forensic trial proceeding.

## 4   An Exemplar

To better comprehend circumstantial-evidence-based effort judgment, we employ one Web service composition project as an example to explain the judgment process. Considering our work is still in the early stage, here we directly choose five hypotheses of development experience as original circumstantial evidences. In practice, however, development experiences should be justified through the evidence collection approach supplied by EBSE.

### 4.1   Five Circumstantial Evidences

In software engineering, effort of a task is generally accounted by calculating how long and how many workers are needed to finish the task. In other words, the amount of human activities in a project is proportional to the amount of effort required to finish the project. Therefore, for a certain software project, one basic circumstantial evidence (CE) can be:

**CE1.** In general, the increase of required human activities in a project will have a proportional impact on the final effort.

Human activities include both physical and mental activities. Since software development is a knowledge-intensive undertaking, software product/service is a type of intellectual property produced by human mental activities. Unfortunately, within a given time span people have limited mental capability to deal with information [8]. For every single person, the increased amount of information beyond a certain point may even defeat his/her mental ability, and hence result in errors [10]. As a result, the more information that exists in a project, the more people and human activities might be required to perform accurate manipulations. Together with CE1, therefore, we can find a new circumstantial evidence:

**CE2.** In general, the increase of information in a project will have a proportional impact on the final effort.

Based on our common experience, the adoption of sophisticated tools usually implies much information we have to deal with in a project. However, tools are essentially developed and used to save human activities. For a certain project, the more work the tools can fulfill, the less human activities the project will require. Consequently, also together with CE1, a tool-related circumstantial evidence is:

**CE3.** In general, the increase of work that tools can fulfill in a project will have an inversely proportional impact on the final effort.

In the software economics field, complexity has been viewed as an inherent property of the functional requirements of a software product, which cannot be reduced or simplified beyond a certain threshold [19]. Moreover, complexity has been proved to be a significant and non-negligible factor that influences software development and maintenance [11]. In fact, the more complexity involved in a system, the more difficulty the designers or engineers have to understand the implementation process and thus the system itself [9], and hence the greater mental effort people have to exert to solve the complexity [8].

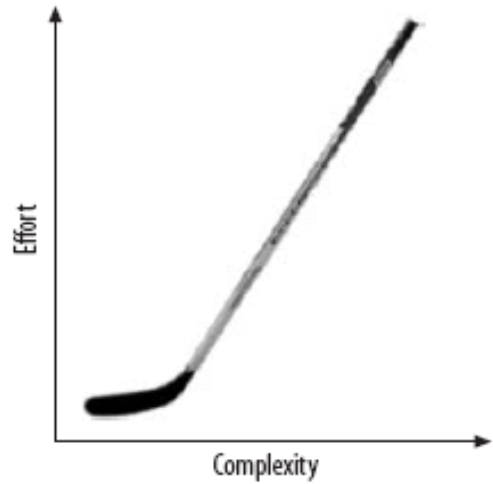

**Fig. 2.** The hockey stick function.

The hockey stick function [13] vividly depicts the relationship between complexity and effort of a software project, as illustrated in Fig. 2. The amount of required effort may suddenly increase when the corresponding project exceeds a certain level of complexity. Overall, the circumstantial evidence related to complexity can be summarized as:

**CE4.** In general, the increase of complexity in a project will have a proportional impact on the final effort.

When it comes to project complexity, one of the main contributors is the complexity of the methods that target achieving the project goals [12]. The methods of software development are mainly reflected by the techniques used to implement a corresponding project. In particular, techniques have been viewed as internal environment of a system (organization), while the system's complexity is considered a response to the environmental complexity [13]. Consequently, the complexity of techniques involved in a software project will positively influence the complexity of the project. Therefore, together with CE4, we can identify the circumstantial evidence CE5:

**CE5.** In general, the increase of difficulty of techniques in a project will have a proportional impact on the final effort.

## 4.2 Circumstantial-Evidence-Based Judgment for a Web Service Composition Project

By using an effort-oriented classification matrix [16], existing approaches to Web service composition can be classified according to different type of contexts and techniques. For example, we can distinguish between Orchestration and Choreography in consideration of composition pattern, Syntactic and Semantic compositions according to the semiotic context, or REpresentational State Transfer (RESTful) and Simple Object Access Protocol (SOAP) based compositions according to the composition mechanism. Since different types of Web service composition require different development activities, a part of the profile exploration of a Web service composition project can be done as follows.

*Orchestration vs. Choreography:* Orchestration normally describes and executes a centralized process flow that acts as a coordinator to the involved Web services. The central coordinator explicitly specifies the business logic and controls the order of invocation of Web services. Choreography represents collaboration between web services that focuses on the peer-to-peer message exchange. The collaboration is decentralized where all participating Web services work equally and do not rely on a central controller. Since distributed processing would be inevitably more complicated than non-distributed processing [14], generally speaking, for the same Web service composition project the choreography-based implementation will be more complex than the orchestration-based implementation. Meanwhile, as the current de facto standard of orchestrating Web services, Business Process Execution Language (BPEL) stemmed from existing languages and tools and has been widely accepted, whereas the choreography language Web Services Choreography Description Language (WS-CDL) was developed without any prior implementation and is still far from mature [17]. Considering this technical influence, the implementation of choreography will be more difficult than that of orchestration. Consequently, if holding the other aspects of one particular Web service composition project constant, development actions (DA) can be abstracted and compared between orchestration and choreography:

**DA1.** In general, the implementation of choreography is more complex than that of orchestration.

**DA2.** In general, the techniques used for choreography are more difficult than that for orchestration.

*Syntactic composition vs. Semantic composition:* The syntactic Web, for example the current World Wide Web, was designed primarily for human interpretation and conveying information. The lack of machine-readable semantics then requires human intervention for Web service discovery and composition, and therefore hampers the usage of Web services in complex business environment. On the contrary, the semantic Web and semantic Web service were proposed through incremental and information-added adjustments. Since semantic Web and semantic Web services are supposed to automate service discovery, selection, composition and execution by adding the inherent meanings [18], human activities within semantic compositions will be decreased while the involved information will be increased. However, the increased information in semantic Web service composition is for machine interpretation rather than human intervention. Meanwhile, syntactic and semantic Web services share the unified Web infrastructure and both use markup language based techniques to describe information. It can then be stated that the difficulty levels of techniques adopted in both syntactic and semantic Web service compositions are similar. Therefore, for proposals with different semiotic context for a particular Web service composition project, we can assert:

**DA3.** In general, the implementation within syntactic context requires more human interventions than that within semantic context.

**DA4.** In general, the implementation within semantic context involves more information for machine interpretation than that within syntactic context.

**DA5.** In general, the difficulty of techniques used for syntactic implementation is similar to that for semantic implementation.

*RESTful composition vs. SOAP-based composition:* RESTful Web service composition integrates normally disparate Web resources to create a new application. These resources can be the exposure of pure data or traditional application functionality. SOAP/WS-* based Web service composition is a collection of related, structured activities or tasks that produce a specific service or product for a particular customer. Compared with RESTful compositions, SOAP-based compositions employ more sophisticated techniques including heavyweight protocols, a set of WS-* stack, and more Message Exchange Patterns (MEPs), which can satisfy more QoS requirements while also deal with more information. Therefore, if the requirement of a particular Web service composition project can be satisfied by using either RESTful or SOAP-based approach, we can assert:

**DA6.** In general, the techniques used for the SOAP-based implementation are more difficult than that for the RESTful implementation.

**DA7.** In general, the SOAP-based implementation deals with more information than the RESTful implementation does.

To summarize, DA1~DA7 are analysis results drawn from characteristics of different types of Web service composition projects. These analysis results can be viewed as abstracts of different development actions, and act as possible inference bridges between real development actions and the identified circumstantial evidences. Benefiting from TPL based diagnostic reasoning, therefore, we can conveniently and qualitatively judge the effort of these composition types. For example, the forward implication from (DA1, DA2) to (CE4, CE5) can infer that choreography requires more effort than orchestration does when implementing a particular Web service composition project, or "$E_{Ch} > E_{Or}$" for short. Similarly, we can also give qualitative effort judgment for the other composition types mentioned in this Section, as shown in Table 1. Note that not all the circumstantial evidences are applicable in this case.

**Table 1.** Circumstantial-evidence-based effort judgment for different types of Web service compositions.

|  | Development Actions | Circumstantial Evidences | Effort Judgment |
|---|---|---|---|
| **Orchestration** | (DA1, DA2) | (CE4, CE5) | $E_{Ch} > E_{Or}$ |
| **Choreography** | | | |
| **Syntactic** | DA3 | CE1 | $E_{Sy} > E_{Se}$ |
| **Semantic** | | | |
| **RESTful** | (DA6, DA7) | (CE5, CE2) | $E_{SO} > E_{RE}$ |
| **SOAP-based** | | | |

## 5 Conclusion

Expert judgment is the widely adopted technique for software effort estimation in industry. From an evidence-based perspective, expert judgment relies on direct evidences that require the availability of both experts and past project data. Considering the lack of suitable experts and available data in the current practice of software engineering, we propose circumstantial-evidence-based judgment to facilitate qualitative effort estimate of a new software project. Compared with direct evidences that focus on actual effort of past projects, circumstantial evidences for effort judgment are abstracts of existing software development experiences. Before implementing a new project, identified circumstantial evidences can be combined with the profile of new project by rational inferences to qualitatively compare the efforts of different development proposals. As such, circumstantial-evidence-based judgment can not only help settle implementation design for software project, but also act as complementary to expert judgment for the implementation effort. Moreover, the circumstantial evidences in the context of effort judgment can be accumulated and deposited as general knowledge to further guide and assess individual expert judgments. SLR, as the main methodology applied for EBSE, can be an effective approach to evidence collection for circumstantial-evidence-based judgment. All the development experiences mentioned in this paper do need the support by further evidences that can be identified and synthesized by this EBSE methodology. Therefore, our future work will try to apply SLR to this novel effort judgment method. Moreover, we also plan to use propositional calculus to formalize the rational inference taking place during judgment processes.

## Acknowledgment

NICTA is funded by the Australian Government as represented by the Department of Broadband, Communications and the Digital Economy and the Australian Research Council through the ICT Centre of Excellence program.